\documentclass[aps,twocolumn,showpacs,preprintnumbers,amsmath,amssymb,superscriptaddress,floatfix,nofootinbib]{revtex4}

\usepackage{graphicx}
\usepackage{epsfig}
\usepackage{epstopdf}
\usepackage{hyperref}
\usepackage{amsmath}
\usepackage{amsfonts}
\usepackage{amssymb}

\begin{document}

\title{\boldmath Semileptonic decay of $B^{-}_c$ into $X(3930)$, $X(3940)$, $X(4160)$}
\date{\today}

\author{Natsumi Ikeno} \email{ikeno@rs.tottori-u.ac.jp}
\affiliation{
Department of Life and Environmental Agricultural Sciences,
Tottori University, Tottori 680-8551, Japan}

\author{Melahat Bayar} \email{melahat.bayar@kocaeli.edu.tr}
\affiliation{Department of Physics, Kocaeli University, 41380 Izmit, Turkey}

\author{Eulogio Oset} \email{oset@ific.uv.es}
\affiliation{Departamento de
F\'{\i}sica Te\'orica and IFIC, Centro Mixto Universidad de
Valencia-CSIC Institutos de Investigaci\'on de Paterna, Aptdo.
22085, 46071 Valencia, Spain}

\begin{abstract}
We study the semileptonic decay of $B^{-}_c$ meson into $\bar{\nu}  l^-$ and the isospin zero $X(3930)~(2^{++})$, $X(3940)~(0^{++})$, $X(4160)~(2^{++})$ resonances. We look at the reaction from the perspective that these resonaces appear as dynamically generated from the vector-vector interaction in the charm sector, and couple strongly to $D^{*}\bar{D^{*}}$ and  $D^{*}_{s}\bar{D_{s}^{*}}$. We also look into the  $B^{-}_c \rightarrow \bar{\nu}_{l} l^- D^{*}\bar{D^{*}}$ and $B^{-}_c \rightarrow \bar{\nu}_{l} l^- D^{*}_{s}\bar{D_{s}^{*}}$ reactions close to threshold and relate the $D^{*}\bar{D^{*}}$ and  $D^{*}_{s}\bar{D_{s}^{*}}$ mass distribution to the rate of production of the $X$ resonances. 
\end{abstract}


\maketitle
\section{Introduction}

The $X, Y, Z$ states, that challenge the  
constituent quark model picture of meson~\cite{Godfrey:1985xj,Vijande:2004he}
have been one of the most spectacular findings
in hadron spectroscopy recently~\cite{Chen:2016qju,Liu:2013waa,Godfrey:2008nc,Guo:2017jvc}.
Their advent has  stimulated much theoretical work aimed at unravelling their structure. 
Tetraquark pictures have been proposed~\cite{Chen:2016oma,Esposito:2016noz}
as well as molecular pictures stemming from
the interaction of more elementary mesons~\cite{Ortega:2012rs,Molina:2009ct,Guo:2017jvc}.
One of these pictures deals with the interaction of vector mesons with charm,
leading to hidden charm quasibound meson states~\cite{Molina:2009ct}.
In that work, the channels $D^{*}\bar{D^{*}}$, $D^{*}_{s}\bar{D_{s}^{*}}$,
$K^{*}\bar{K}^{*}$, $\rho \rho$, $\omega \omega$, $\phi \phi$,
$J/\psi J/\psi$, $\omega J/\psi$, $\phi J/\psi$, $\omega \phi$ 
were considered and the interaction between them was 
obtained using an extention of the local hidden gauge approach~\cite{Bando:1987br,Harada:2003jx,Meissner:1987ge},
exchanging vector mesons, and through contact terms provided by the theory. 
Some quasibound states were found which could be associated to known resonances. 
These states were: one state around $3943$ MeV with $I^G [J^{PC}] = 0^{+} [0^{++}]$,
which was associated to the
$X(3940)$~\cite{Abe:2004zs,Abe:2007jna};
another state around 3922~MeV with $0^{+} [2^{++}]$,
which was associated to the $X(3930)$~\cite{Uehara:2005qd}
 (now classified in the PDG~\cite{Patrignani:2016xqp} as the the $\chi_{c2} (2P)$ ),
which could also correspond to the $X(3915)$~\cite{Zhou:2015uva,Ortega:2017qmg},
and a third one at $4169$ MeV  with $0^{+}[2^{++}]$ that was associated to the
$X(4160)$~\cite{Abe:2007sya}.

The $X(3940)$ was found to couple mostly to  $D^{*}\bar{D^{*}}$ in ~\cite{Molina:2009ct}, the $X(3930)$ also had its strongest  coupling to $D^{*}\bar{D^{*}}$ and  the $X(4160)$ had its strongest coupling to  $D^{*}_{s}\bar{D_{s}^{*}}$. The light vector-vector channels couple weakly to those states, but given the large space available, they are the biggest source of the width, which in the theoretical work is also found in reasonable agreement with experiment. It is interesting to mention that there is a large list of works suggesting a bound  $D^{*}_{s}\bar{D_{s}^{*}}$ state \cite{Liu:2009ei,Branz:2009yt, Weinberg:1965zz, Baru:2003qq, Chen:2015fdn, Karliner:2016ith}. QCD sum rules,  although with its usual large uncertainties,  have also speculated on this possibility \cite{Albuquerque:2009ak, Zhang:2009st}. The curious thing is that all these work aimed at reproducing the $X(4140)$ resonance not the $X(4160)$. One can think that the fact that light vector channels were not included as coupled channel in these studies had as a consequence a small width for the resonance which made it more appealing to have it associated to the $X(4140)$. Yet, the quantum numbers $0^{+}[1^{++}]$ for this resonance determined lately make the association of the state found in these works to the $X(4140)$ inappropriate, while the association to the $X(4160)$ is more natural. 

The discussion on these states becomes more actual when one recalls the experimental work \cite{Aaij:2016iza, Aaij:2016nsc} in the $B^+\to J/\psi \phi K^+$ reaction, where the data analysis brought the surprising result that the $X(4140)$ has a width of around $ \Gamma \simeq 83$ MeV, while former experiments give a width around $19$ MeV  \cite{Aaltonen:2009tz, Brodzicka:2010zz, Aaltonen:2011at, Aaij:2012pz, Chatrchyan:2013dma, Abazov:2013xda, Lees:2014lra, Abazov:2015sxa}. This puzzle found a reasonable explanation in a recent work \cite{Wang:2017mrt}, where the $ J/\psi \phi $ invariant mass distribution at low invariant masses was analyzed in terms of the $X(4140)$ and $X(4160)$ and the mass distribution was better reproduced. 

The result of this new analysis was that the $X(4140)$ has a width compatible with $19$ MeV, and it is the $X(4160)$ resonance the one that fills the strength in that region. The striking thing is that, since the $X(4160)$ in that work is supposed to be a $D^{*}_{s}\bar{D_{s}^{*}}$ bound state, but with a relatively large coupling to $ J/\psi \phi $ in the coupled channels study of \cite{Molina:2009ct}, when the  $ J/\psi \phi $ mass distribution is studied, a large cusp structure develops in this distributions at the  $D^{*}_{s}\bar{D_{s}^{*}}$ threshold and such cusp is present in the experiment. 

In view of this puzzling situation, any other reaction that brings light into these issues should be most welcome. This is the purpose of the present work, where we propose to measure the semileptonic decay of $B^{-}_c \rightarrow \bar{\nu}_{l} l^- X_{i}$, with $ X_{i} $ any of the three resonances, $X(3930)$, $X(3940)$, $X(4160)$.  Actually, this reaction has been studied recently \cite{wangzang} from the perspective that the $X(3940)$ and $X(4160)$ resonances are radial high excitations of the charmonium states, corresponding to the $ \eta_{c} (3S) $ and $ \eta_{c} (4S) $ respectively. Our picture, where these resonances are generated from the interaction of vector meson with charm is quite different and the study of the $B_c \rightarrow \nu_{e} e^+ X_{i}$ decays from this perspective is worth pursuing. 

The use of semileptonic weak decays aiming at determining  the structure of resonances has been exploited before in different cases. In \cite{Navarra:2015iea} the $ B_{s} $ and $ B $ semileptonic decays,  $\bar{B}^{0}_{s}\rightarrow D^{*}_{s0} (2317) \bar{\nu}_{l} l^- $, $\bar{B}^{0}\rightarrow D^{*}_{0}(2400)^{+} \bar{\nu}_{l} l^- $ were studied and compared to related reactions like  $\bar{B}^{0}_{s}\rightarrow (DK)^{+} \bar{\nu}_{l} l^- $. In \cite{Sekihara:2015iha} the production of light scalar mesons and light vector mesons was also investigated in semileptonic decays of $ D $ and $ D_{s} $ mesons. In \cite{Ikeno:2015xea} the $ \Lambda_{c} \rightarrow \Lambda(1405) e^{+} \nu_{e} $ was studied, looking at the decays of $ \Lambda(1405) $ into $ \pi^{+}  \Sigma^{-}$, $ \pi^{-}  \Sigma^{+}$, $ \pi^{0}  \Sigma^{0}$ and  $ \bar{K} N$ production. In this case the weak decay filters $ I=0 $ in the final meson baryon system, which makes this reaction special to investigate the properties of the $ \Lambda(1405) $. Related to these works, but with a different aim, one has the work of \cite{Liang:2016exm} where the $ \Lambda_{b} \rightarrow \Lambda_{c}(2595)  \bar{\nu}_{l} l $ and $ \Lambda_{b} \rightarrow \Lambda_{c}(2625)  \bar{\nu}_{l} l $ decays are studied in order to test the pseudoscalar-baryon and vector-baryon components of the $  \Lambda_{c}(2595) $ and $ \Lambda_{c}(2625) $ resonances \cite{Liang:2014kra}. 

In the present work we take advantage of these previous studies and calculate the $B_c \rightarrow \nu_{l} l^+ X_{i}$ decay rates and compare them with the $B_c \rightarrow \nu_{l} l^+ D^{*}\bar{D^{*}}$, $\nu_{l} l^+ D^{*}_{s}\bar{D_{s}^{*}}$ decays. We establish a link between these processes which is tied to the molecular nature of these resonances, making predictions to be tested in future experiments from where much valuable information concerning the nature of these states is to be expected. 

\section{Formalism}
\begin{figure}[tb]
\begin{center}
\includegraphics[height=3cm]{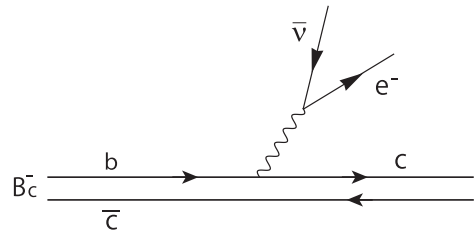}
\caption{Diagrammatic representation of the quark level for $B_c
 \rightarrow \nu_{e} e^- (c \bar{c})$.}
\label{fig:1}
\end{center}
\end{figure}

The $B^{-}_c \rightarrow \bar{\nu} e^- X$ process proceeds at the
quark level through a first step shown in Fig.~\ref{fig:1}.
The process involves the $bc$ weak transition, which is the same one as in the $B$ decays studied in Ref.~\cite{Navarra:2015iea}.

\begin{figure}[tb]
\begin{center}
\includegraphics[height=3.5cm]{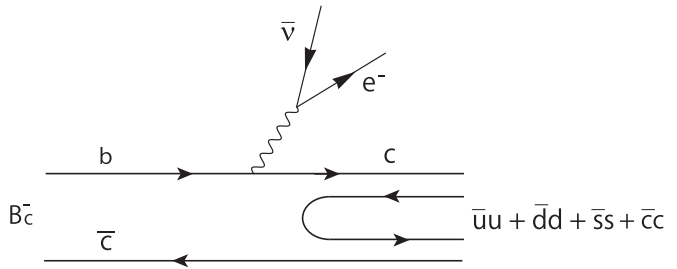}
\caption{Dominant mechanism for the hadronization into two mesons
 of the $c \bar{c}$ state after the weak process.}
\label{fig:2}
\end{center}
\end{figure}

There is, however, a novelty in the present process.
Indeed, if we want to see two mesons,
the $c \bar{c}$ quarks of Fig.~\ref{fig:1} must hadronize into two mesons components.
This is easily done  for mesons 
since one introduces an extra $\bar{q} q$ pair
with vacuum quantum number, $\bar{u}u + \bar{d}d + \bar{s}s +  \bar{c}c$, and then
the two quarks after the weak process participate
in the formation of the two mesons.
With two quarks after the weak vertex, as in Fig.~\ref{fig:2}, the new
$\bar{q} q$ pair can be placed in between these $c\bar{c}$
quarks.

The procedure followed here is inspired in the approach of 
Ref.~\cite{Liang:2015twa} 
where the basic mechanisms at the quark level are
investigated, 
then pairs of hadrons are produced after implementing
hadronization, and finally these hadrons are allowed to undergo final
state interaction.

The hadronization of $c \bar{c}$, introducing the $q \bar{q}$ pair is done as follows ~ \cite{Liang:2015twa, Navarra:2015iea}. We take the $q \bar{q}$ matrix $M$, 
\begin{eqnarray}
M \equiv q \bar{q}^{\tau} = \left( 
\begin{array}{cccc}
 u\bar{u} & u\bar{d} & u\bar{s} & u\bar{c} \\
 d\bar{u} & d\bar{d} & d\bar{s} & d\bar{c} \\
 s\bar{u} & s\bar{d} & s\bar{s} & s\bar{c} \\
 c\bar{u} & c\bar{d} & c\bar{s} & c\bar{c}  
\end{array}
\right)
\end{eqnarray}

Then 
\begin{eqnarray}
 c(\bar{u}u + \bar{d}d + \bar{s}s + \bar{c}c ) \bar{c}= \sum_{i=1}^{4} M_{4i} M_{i4} = (M^2)_{44}
\end{eqnarray}

We now write $M$ in terms of vector mesons and we have the vector matrix $V$,
\begin{eqnarray}
M \rightarrow
V \equiv \left( 
\begin{array}{cccc}
\frac{\rho^0}{\sqrt{2}} + \frac{\omega}{\sqrt{2}} & \rho^+ & K^{* +} & \bar{D}^{*0} \\
\rho^- & -\frac{\rho^0}{\sqrt{2}}+\frac{\omega}{\sqrt{2}}  & K^{* 0} & \bar{D}^{*-} \\
K^{*-} & \bar{K}^{*0} & \phi & \bar{D}_{s}^{*-} \\ 
D^{*0} & D^{*+} & D_{s}^{*+} & J/ \psi
\end{array}
\right).
\label{eq:V}
\end{eqnarray}

Then $ M^{2}  $ becomes $ V^{2} $ and 

\begin{equation}
 (V \cdot V)_{44} = D^{*0} \bar{D}^{*0} + D^{*+} \bar{D}^{*-} + D^{*+}_{s} \bar{D}^{*-}_{s}
  + J/\psi J/\psi .
\end{equation}

We neglect the $J/\psi J/\psi$ channel since it has too high
energy relative to  other channels. 
Only an $I=0$ state is produced from the $c \bar{c} $ component 
since the hadronization is a strong interaction and does not change isospin.
We can write the $D^{*} \bar{D}^{*} $ combination in terms of 
the isospin doublets $(D^{*+}, -D^{*0})$ and $(\bar{D}^{*0}, \bar{D}^{*-})$ and then
the production vertex is written as~\cite{Liang:2015twa}
\begin{equation}
 (V \cdot V)_{44} \rightarrow \sqrt{2} | D^{*} \bar{D}^{*} ; I=0 \rangle 
+  | D^{*}_{s} \bar{D}^{*}_{s} ; I=0 \rangle.
\end{equation}
 
\begin{figure}[tb]
\begin{center}
\includegraphics[width=8cm]{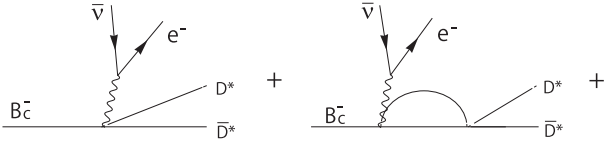}
\caption{Diagrams involved in the final state interaction of the primary
 $D^{*} \bar{D^{*}}$ mesons, (a) tree level, (b) rescattering.}
\label{fig:3}
\end{center}
\end{figure}

The final state interaction of $D^{*} \bar{D}^{*}$ is depicted in Fig.~\ref{fig:3},
and 
we use the interaction of Ref.~\cite{Molina:2009ct} 
where, using an extension of the local hidden gauge approach,
the interaction of $D^{*} \bar{D}^{*}$ generates several resonances, and
some $XYZ$ states were dynamically generated.
As suggested in Refs.~\cite{Liang:2015twa,Molina:2009ct},
the resonances most strongly coupled to the $D^{*} \bar{D}^{*}$ channel  
correspond to the experimental states $Y(3940), Z(3930)$
and the $D^{*}_{s} \bar{D}^{*}_{s}$ channel corresponds to 
the $X(4160)$.

\subsection{ Coalescence}
\begin{figure}[tb]
\begin{center}
\includegraphics[height=3cm]{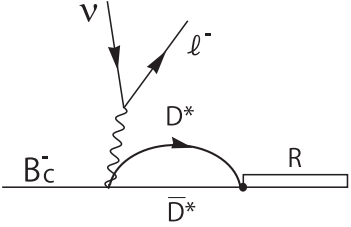}
\caption{Diagrams of the coalescence process which produce the resonances 
after rescattering.}
\label{fig:4}
\end{center}
\end{figure}

From the present perspective, we have studied the semileptonic decay
process in Ref. \cite{Navarra:2015iea}.
First, we study the coalescence process which produces the resonances 
after rescattering, as shown in Fig.~\ref{fig:4}.

%
This process has a three-body final state with a lepton, its
neutrino and the resonance $R$. 
The resonance $R$ stands for the $X,Y,Z$ resonances.
The hadronization factor $V_{\rm had}$ can be obtained as
\begin{eqnarray}
 V_{\rm had} = C (\sqrt{2} \  G_{D^{*} \bar{D}^{*}} 
 \ g_{R, D^{*} \bar{D}^{*}} 
+   G_{D^{*}_{s} \bar{D}^{*}_{s}}  \ g_{R, D^{*}_{s} \bar{D}^{*}_{s}}),
\label{eq:Vhad_coa}
\end{eqnarray}
for the resonance $R$ in $J=0$ which requires $L=0$.
For the constant $C$, we use the value $C=7.22$
of the semileptonic $B$ decays 
as established in Ref.~\cite{Navarra:2015iea}.
$G_{D^{*} \bar{D}^{*}}$ and $ G_{D^{*}_{s} \bar{D}^{*}_{s}}$ are 
the two meson loop functions, 
and
$g_{R, D^{*} \bar{D}^{*}} $ and $ g_{R, D^{*}_{s} \bar{D}^{*}_{s}}$ are
the couplings of the resonance to these channels.
We use the values reported in Ref.~\cite{Molina:2009ct}.
%
%
The two meson loop function $G_i$ for each channel $i$ is
\begin{eqnarray}
G_i (s) = i \int \frac{d^4 q}{(2 \pi)^4} 
\frac{1}{q^2 -m_1^2 + i\epsilon}
\frac{1}{(P-q)^2 -m_2^2 + i\epsilon},
\label{eq:Gi}
\end{eqnarray}
where $P$ is the total four-momentum of the two mesons, and
$m_1$ and $m_2$ are the masses of the two mesons in channel $i$.
We use cut off regularization as done in \cite{Wang:2017mrt} to avoid potential problems of dimensional regularization pointed out in \cite{Wu:2010rv}. The $G$ function has the form

\begin{eqnarray}
 G_i  =  \int^{q_{max}}_{0} \frac{q^{2} ~dq}{(2\pi)^2} \dfrac{\omega_{1} + \omega_{2} }{\omega_{1} ~ \omega_{2} \Big( (P^{0})^{2} -(\omega_{1} + \omega_{2})^{2} +i~\epsilon \Big )} ,
 \label{eq:Gi2}
\end{eqnarray}
where $q_{max}  $ stands for the cutoff in the three momentum, the square of center of mass energy $ (P^{0})^{2}=s $ and  $ \omega_{i} = \sqrt{\vec{q}^{~2}_{i}+m^{2}_{i}} $. As in   \cite{Wang:2017mrt} we use $q_{max} =690 $ MeV.

The whole amplitude $T_{B_c}$ for the semileptonic decay of the $B_c$ meson is written as,
\begin{equation}
 T_{B_c} = -i \frac{G_{\rm F}
  V_{bc}}{\sqrt{2}}L^{\alpha}Q_{\alpha} V_{\rm had},
\end{equation}
where
\begin{equation}
 L^\alpha = \bar{u}_\nu \gamma^\alpha (1 - \gamma_5) v_l, 
\hspace{3mm}
 Q_\alpha = \bar{u}_c \gamma_\alpha (1 - \gamma_5) u_b.
\end{equation}
By following the steps of Refs.~\cite{Navarra:2015iea,Sekihara:2015iha}, 
we find for the sum
and average over the polarization of the fermions
\begin{eqnarray}
 \frac{1}{2} \sum_{\rm pol} |T_{B_c}|^2 = 
\frac{ 4 |G_{\rm F} V_{bc} V_{\rm had} |^2}
{m_e m_\nu m_{B_c} m_R}
(p_{B_c} \cdot p_\nu)(p_R \cdot p_e).
\end{eqnarray}
Further steps are done in Ref.~\cite{Navarra:2015iea} to perform the angular
integrations of the resonance in the $B_c$ rest frame and 
the lepton in the $\nu e$ rest frame, 
and 
finally one obtains a formula of the decay widths $\Gamma_{\rm coal} $
for the coalescence of the $X,Y,Z$ resonances by
\begin{eqnarray}
 \Gamma_{\rm coal}  &=& \frac{|G_{\rm F} V_{bc} V_{{\rm had}}|^2}
{8 \pi^3 m_{B_c}^3 m_{R} } \nonumber\\
&\times&
\int dM_{\rm inv}^{(\nu e)} P^{\rm cm}_{R} \tilde{p}_{\nu} 
| M_{\rm inv}^{(\nu e)} |^2 \left( \tilde{E}_{B_c} \tilde{E}_{R} - \frac{\tilde{p}_{B_c}^2}{3} \right).
\nonumber\\
\label{eq:dGam}
\end{eqnarray}
Here, $P^{\rm cm}_{R}$ is the momentum of the resonance $R$ in the $B_c$
rest frame, and
 $\tilde{p}_{\nu}$ is the momentum of the neutrino in the
$\nu e$ rest frame,
\begin{eqnarray}
P^{\rm cm}_{R} &=& \frac{\lambda^{1/2}(m_{B_c}^2, [M_{\rm inv}^{(\nu e)}]^2,
  m_{R}^{2} )}{2m_{B_c} },  
\\
\tilde{p}_{\nu} &=& \frac{\lambda^{1/2}([M_{\rm inv}^{(\nu e)}]^2, m_{\nu}^2,
  m_{e}^2 )}{2M_{\rm inv}^{(\nu e)} },
\label{eq:p_nu}  
\end{eqnarray}
with the K\"{a}llen function defined as, 
\begin{equation}
 \lambda(a, b, c)=a^2+b^2+c^2-2ab-2bc-2ca.
\label{kallen}
\end{equation}
The energies $\tilde{E}_{B_c}$ and $\tilde{E}_R$  
are calculated in the $\nu e$ rest frame,
\begin{eqnarray}
\tilde{E}_{B_c} &=& \frac{m_{B_c}^2 + [M_{\rm inv}^{(\nu e)}]^2 -
 m_{R}^{2} }{2 M_{\rm inv}^{(\nu e)} }, \\
\tilde{E}_{R} &=& \frac{m_{B_c}^2 - [M_{\rm inv}^{(\nu e)}]^2 -
 m_{R}^{2} }{2 M_{\rm inv}^{(\nu e)} },
\end{eqnarray}
and $\tilde{p}_{B_c}$ is the momentum of the $B_c$ in the
$\nu e$ rest frame,
\begin{eqnarray}
\tilde{p}_{B_c}^{2} &=& \tilde{E}_{B_c}^2 - m_{B_c}^2.
\label{eq:p_Bc} 
\end{eqnarray}
The integral range of $M_{\rm inv}^{(\nu e)}$ is
$ [m_e + m_\nu, m_{B_c} - m_R]$.
We take the Fermi coupling constant
$G_{\rm F} = 1.166 \times 10^{-5}$ GeV$^{-2}$ and
the Cabibbo-Kobayashi-Maskawa matrix element 
$V_{bc} = 0.0411$.

For the other resonance of $2^{++}$, we can not use the formula in Eq.~(\ref{eq:dGam})
because we need $L=2$ state and hence the matrix element would be
different. For the $L=2$ case we replace the $ P^{\rm cm}_{R} $ with $ (P^{\rm cm}_{R})^{5} $ in Eq. \ref{eq:dGam}, but we do not obtained an absolute value for the width. Yet, we can obtain the ratio of rates for the two $ 2^{++} $ resonances. 




\vspace{1cm}

\subsection{Rescattering}
Next, we study the rescattering in the final states $D^{*} \bar{D^{*}}$ and  
$D_{s}^{*} \bar{D_{s}^{*}}$  as shown in Fig.~\ref{fig:3}.
The different final states of the $D^{*} \bar{D^{*}}$ and  
$D_{s}^{*} \bar{D_{s}^{*}}$ 
are treated separately.
%
%
The amplitude $V_{\rm had}^{\prime} $ 
in the $D^{*} \bar{D^{*}}$ states for $I=0, ~J=0$  and $I=0,~ J=2$,  
where the resonances couple strongly to $D^{*} \bar{D^{*}}$,
is written as,
\begin{eqnarray}
 V_{\rm had}^{\prime} &=& C (\sqrt{2} 
+ \sqrt{2} \  G_{D^{*} \bar{D}^{*}} 
 \ t_{ D^{*} \bar{D}^{*},  D^{*} \bar{D}^{*} }  \nonumber\\
& & + \ G_{D_{s}^{*} \bar{D}_{s}^{*}} 
 \ t_{ D^{*}_{s} \bar{D}^{*}_{s}, D^{*} \bar{D}^{*} }).
\end{eqnarray}
where $G$ is the two meson loop function for each channel in
Eq.~(\ref{eq:Gi2}). The factor $ C $ is not the same for $ J=0 $ and $ J=2 $, and we will come back to that.

On the other hand, 
in the other case of resonance coupled to the $D_{s}^{*} \bar{D_{s}^{*}}$ state for $I=0, J=2$,
the hadronization amplitude $V'_{\rm had}$ is written as,
\begin{eqnarray}
 V_{\rm had}^{\prime} &=& C' ( 1 + \sqrt{2} \  G_{D^{*} \bar{D}^{*}} 
 \ t_{ D^{*} \bar{D}^{*}, D^{*}_{s} \bar{D}^{*}_{s} }  \nonumber\\
& &+  \  G_{D_{s}^{*} \bar{D}_{s}^{*}} 
 \ t_{ D^{*}_{s} \bar{D}^{*}_{s},  D^{*}_{s} \bar{D}^{*}_{s} }).
\end{eqnarray}

The scattering amplitudes $t_{ D^{*} \bar{D}^{*},  D^{*} \bar{D}^{*} } $, 
$ t_{ D^{*}_{s} \bar{D}^{*}_{s},D^{*} \bar{D}^{*} } $ and
$t_{ D^{*}_{s} \bar{D}^{*}_{s}, D^{*}_{s} \bar{D}^{*}_{s} }$
for  the
$D^{*} \bar{D}^{*} \rightarrow  D^{*} \bar{D}^{*}$,
$D^{*} \bar{D}^{*} \rightarrow  D^{*}_{s} \bar{D}^{*}_{s}$, and
$D^{*}_{s} \bar{D}^{*}_{s} \rightarrow  D^{*}_{s} \bar{D}^{*}_{s}$
transitions are written as
\begin{eqnarray}
  t_{ D^{*} \bar{D}^{*},  D^{*} \bar{D}^{*} } 
&=& \frac{g_{R,D^{*} \bar{D}^{*}} g_{R,D^{*} \bar{D}^{*}} }
 {M^2_{\rm inv} - M_{R}^2 + i M_{R} \Gamma_{R}   }, 
 \label{eq:ampli1}
\end{eqnarray}

\begin{eqnarray}
  t_{ D^{*}_{s} \bar{D}^{*}_{s},D^{*} \bar{D}^{*} } 
&=& \frac{g_{R,D^{*}_{s} \bar{D}^{*}_{s} } g_{R,D^{*} \bar{D}^{*} } }
 {M_{\rm inv}^2 - M_{R}^2 + i M_{R} \Gamma_{R}   }, 
 \label{eq:ampli2}
\end{eqnarray}

\begin{eqnarray}
  t_{ D^{*}_{s} \bar{D}^{*}_{s}, D^{*}_{s} \bar{D}^{*}_{s} } 
&=& \frac{g_{R,D^{*}_{s} \bar{D}^{*}_{s}} g_{R,D^{*}_{s} \bar{D}^{*}_{s}} }
 {M_{\rm inv}^2 - M_{R}^2 + i M_{R} \Gamma_{R} ,  }
 \label{eq:ampli3} 
\end{eqnarray}
where $M_{\rm inv}$ in the denominator refers to the final $D^{*} \bar{D}^{*}$
or $D^{*}_{s} \bar{D}^{*}_{s}$ states.
The scattering amplitude $ t_{ D^{*}_{s} \bar{D}^{*}_{s},D^{*} \bar{D}^{*} } $
is the same as the $ t_{D^{*} \bar{D}^{*}, D^{*}_{s} \bar{D}^{*}_{s} } $ one.
The coupling constants $g_{i}$ are the same as in Eq.~(\ref{eq:Vhad_coa}).


The differential decay widths $\displaystyle \frac{d\Gamma}{dM_{\rm inv}}$ 
are given by~\cite{Navarra:2015iea,Sekihara:2015iha,Ikeno:2015xea} 
\begin{eqnarray}
 \frac{d \Gamma_{i}}{dM_{\rm inv}} &=& \frac{|G_{\rm F} V_{bc} V'_{{\rm had},i}|^2}
{32 \pi^5 m_{B_c}^3 M_{\rm inv}^{(i)} } \nonumber\\
&\times&
\int dM_{\rm inv}^{(\nu e)} P^{\rm cm} \tilde{p}_{\nu} \tilde{p}_{i}
 M_{\rm inv}^{(\nu e)^2} \left( \tilde{E}_{B_c} \tilde{E}_{i} - \frac{\tilde{p}_{B_c}^2}{3} \right) 
\nonumber\\
\label{eq:dGamR}
\end{eqnarray}
where $i$ corresponds to the $D^{*} \bar{D}^{*}$ or $D^{*}_{s} \bar{D}^{*}_{s}$ states.
In the case of the $L=2$ state, we replace $P^{\rm cm}$ by $(P^{\rm cm})^ 5$ as done before in the
integrand of Eq.~(\ref{eq:dGam}) for the coalescence case. 
Here, $P^{\rm cm}$ is the momentum of the $\nu e$ system in the $B_c$
rest frame,
$\tilde{p}_{\nu}$ is the momentum of the neutrino in the
$\nu e$ rest frame as defined in Eq.~(\ref{eq:p_nu}),
and
$\tilde{p}_{i}$ is the relative momentum of the
final mesons in their rest frame, 
\begin{eqnarray}
P^{\rm cm} &=& \frac{\lambda^{1/2}(m_{B_c}^2, [M_{\rm inv}^{(\nu e)}]^2,
 [M_{\rm inv}^{(i)}]^2 )}{2m_{B_c} },  
\end{eqnarray}
\begin{eqnarray}
\tilde{p}_{i} &=& \frac{\lambda^{1/2}([M_{\rm inv}^{(i)}]^2, m_{i}^2,
  m_{i}^{\prime 2} )}{2M_{\rm inv}^{(i)} }, 
\end{eqnarray}
with $m_i$, $m_i^{\prime}$ the two meson masses of the final state.
The energies $\tilde{E}_{B_c}$ and $\tilde{E}_i$ 
are calculated in the $\nu e$ rest frame,
\begin{eqnarray}
\tilde{E}_{B_c} &=& \frac{m_{B_c}^2 + [M_{\rm inv}^{(\nu e)}]^2 -
 [M_{\rm inv}^{(i)}]^2 }{2 M_{\rm inv}^{(\nu e)} },\\
\tilde{E}_{i} &=& \frac{m_{B_c}^2 - [M_{\rm inv}^{(\nu e)}]^2 -
 [M_{\rm inv}^{(i)}]^2 }{2 M_{\rm inv}^{(\nu e)} }.
\end{eqnarray} 
and $ \tilde{p}_{B_{c}} $ is given by Eq. (\ref{eq:p_Bc}).

\section{Results}

\begin{figure}[tb]
\begin{center}
\includegraphics[width=8.5cm]{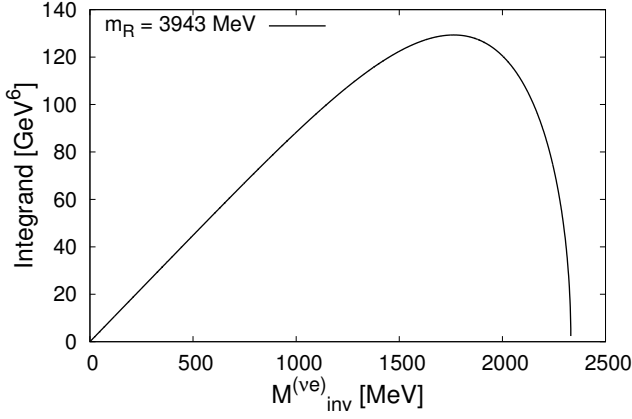}
\caption{The integrand of the integral that appears in
Eq.~(\ref{eq:dGam}) as a function $M_{\rm inv}^{(\nu e)}$ for the
 resonance mass $X(3940)$ of $M_R =3943$~MeV.
}
\label{Integrand_mR3943}
\end{center}
\end{figure}

First, we show the results of the coalescence process.
We consider $X(3934)$ 
as the resonance $R$, namely the $B^{-}_c \rightarrow X(3934) \bar{\nu} l^-$ process.
We evaluate  the decay widths $\Gamma_{\rm coal} $ in Eq.(\ref{eq:dGam})
for the resonance $L=0$ and $J=0$ state
using the mass $m_R = 3943$~MeV, and find
\begin{equation}
 \Gamma_{\rm coal} =  2.6 \times 10^{-14}~{\rm MeV}.
\end{equation}
In Fig.~\ref{Integrand_mR3943} we show the integrand of Eq. (\ref{eq:dGam}). 
The mean life of the $B_c$ is $0.507 \times  10^{-12}$~s, and then 
the branching ratio is evaluated as
\begin{equation}
 \frac{\Gamma_{\rm coal}}{\Gamma_{\rm tot}(B_c)} = 2.0 \times 10^{-5}.
\end{equation}

For the other resonance of $2^{++}$, we can not use the formula 
of the decay widths in Eq.~(\ref{eq:dGam})
because we need an $L=2$ state and hence the matrix element would be
different.
Thus, we calculated the ratio of the two $ L=2 $. We consider two options. In the first one the ratio is evaluated as, 
\begin{equation}
 \frac{\Gamma_{\rm coal}(2)}{\Gamma_{\rm coal}(2')}
= \frac{|V_{\rm had}(2)|^2 /m_R(2)}{|V_{\rm had}(2')|^2 /m_R(2^{\prime})}
 = 1.6
\end{equation}
where $(2)$ indicates the $X(3930)$ resonance state of the 
mass $m_R = 3922$~MeV,
and $(2)^{\prime}$ the $X(4160)$ resonance state of the 
$m_R = 4169$~MeV.

\begin{figure}[htb]
\begin{center}
\includegraphics[width=8.5cm]{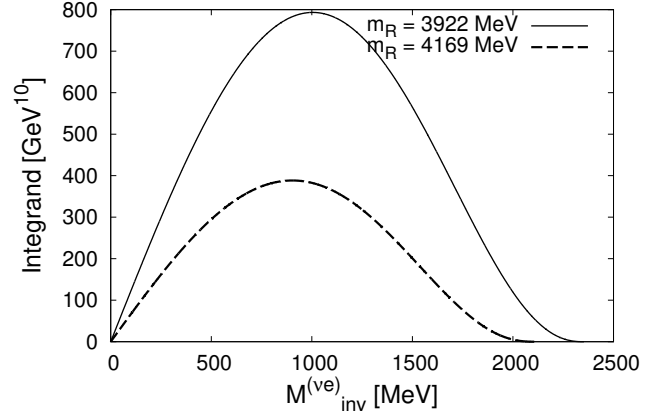}
\caption{The integrands of the integral that appears in
Eq.~(\ref{eq:dGam_L2}) as a function $M_{\rm inv}^{(\nu e)}$ for the
 resonance mass $X(3930)$ of $M_R =3922$~MeV and the resonance mass
 $X(4160)$ of $M_R=4169$~MeV.
}
\label{Integrand_L2}
\end{center}
\end{figure}

For the other option,
we consider that the integral in Eq.~(\ref{eq:dGam}) would have a different form to account for
$L=2$,
\begin{eqnarray}
\int dM_{\rm inv}^{(\nu e)} (P^{\rm cm}_{R})^{5} \tilde{p}_{\nu} 
| M_{\rm inv}^{(\nu e)} |^2 \left( \tilde{E}_{B_c} \tilde{E}_{R} - \frac{\tilde{p}_{B_c}^2}{3} \right).
\label{eq:dGam_L2}
\end{eqnarray}
By replacing the integral in Eq.~(\ref{eq:dGam}) by Eq.~(\ref{eq:dGam_L2}),
the other ratio is evaluated as 
\begin{equation}
 \frac{\Gamma_{\rm coal}(2)}{\Gamma_{\rm coal}(2')}
= 3.6
\end{equation}
This latter result is more realistic and we take it. 
We also show the integrand of Eq.~(\ref{eq:dGam_L2}) in 
Fig.~\ref{Integrand_L2} for the two tensor resonances. 

\begin{center}
  \begin{figure}
  \vspace{10 mm}
   \centering
  \resizebox{0.5\textwidth}{!}{
  \includegraphics{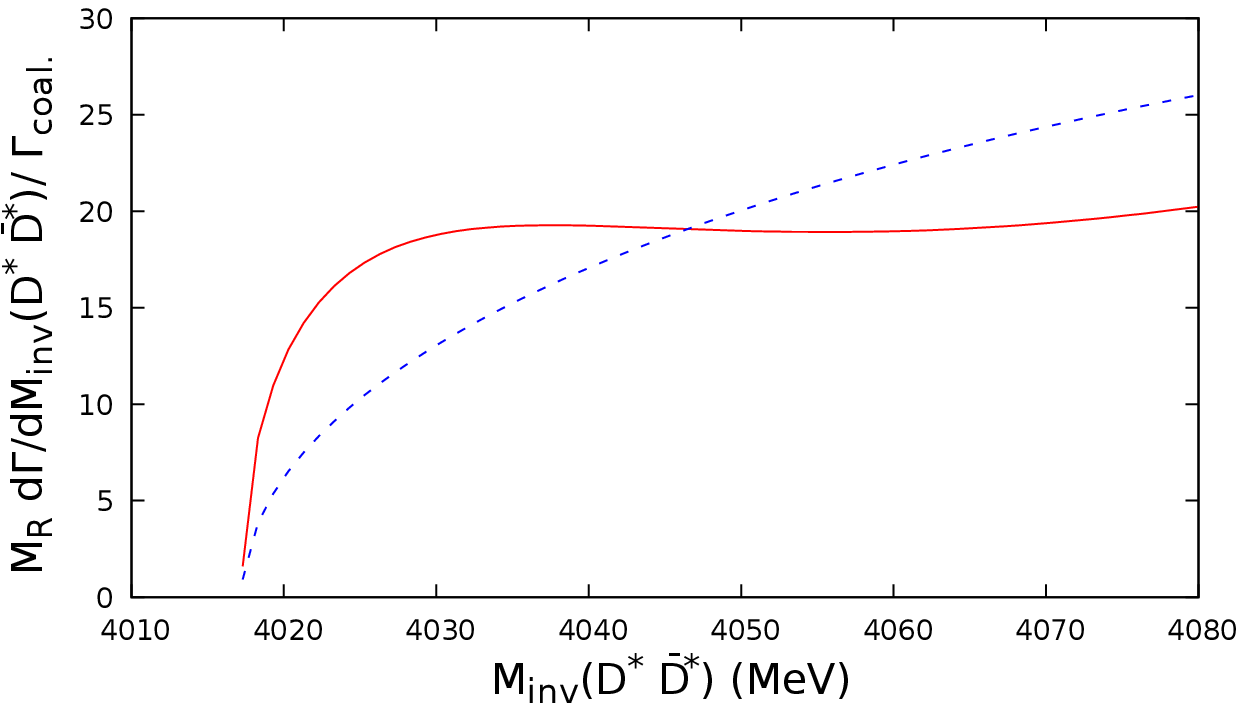}  }
  \caption{ The  $   \frac{M_{R}}{\Gamma_{\rm coal}}\frac{d \Gamma_{i}}{dM_{\rm inv}}  $ as a function of $ M_{inv} (D^{*} \bar{D}^{*})$ for the $B_c \rightarrow \nu_{l} l^+ Y(3940)$  decay. The dashed line corresponds to phase space.}
   \vspace{10 mm}
  \label{fig:J0}
  \end{figure}
  \end{center}

Next, we show the results for the rescattering process. In order to study the rescattering process the scattering amplitudes  $t_{ D^{*} \bar{D}^{*},  D^{*} \bar{D}^{*} } $, 
$ t_{ D^{*}_{s} \bar{D}^{*}_{s},D^{*} \bar{D}^{*} } $ and
$t_{ D^{*}_{s} \bar{D}^{*}_{s}, D^{*}_{s} \bar{D}^{*}_{s} }$  are needed. We use the amplitudes of Eqs. (\ref{eq:ampli1}), (\ref{eq:ampli2}) and (\ref{eq:ampli3}). In Fig. \ref{fig:J0}, using Eqs. (\ref{eq:dGam}) and (\ref{eq:dGamR}), we show the result for $\frac{M_{R}}{\Gamma_{\rm coal}}\frac{d \Gamma_{i}}{dM_{\rm inv}}  $ as a function of $ M_{inv} (D^{*} \bar{D}^{*})$ for the $B^{-}_c \rightarrow \bar{\nu}_{l} l^- X(3940)$  decay, where the dashed line corresponds to a phase space distribution which we normalize to the same area in the range of the figure.  As we can see from Fig. \ref{fig:J0}, the shape of $ D^{*} \bar{D}^{*} $ invariant mass  distribution is different from the phase space.



\begin{center}
  \begin{figure}
  \vspace{10 mm}
   \centering
  \resizebox{0.5\textwidth}{!}{
  \includegraphics{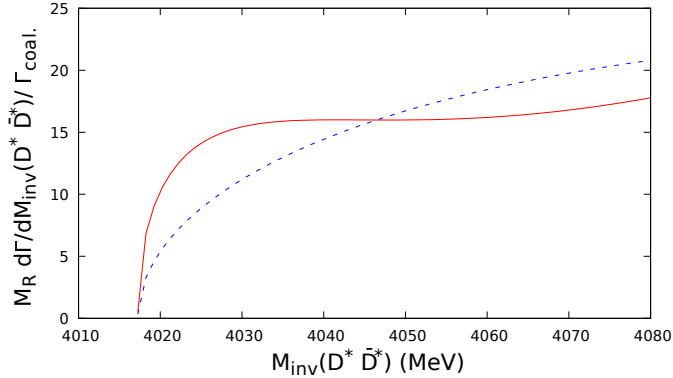}  }
  \caption{The  $   \frac{M_{R}}{\Gamma_{\rm coal}}\frac{d \Gamma_{i}}{dM_{\rm inv}}  $ as a function of $ M_{inv} (D^{*} \bar{D}^{*})$ for the $B_c \rightarrow \nu_{l} l^+ Z(3930)$  decay. The dashed line corresponds to  phase space.}
   \vspace{10 mm}
  \label{fig:J2}
  \end{figure}
  \end{center}

  We have also evaluated  $\frac{M_{R}}{\Gamma_{\rm coal}}\frac{d \Gamma_{i}}{dM_{\rm inv}}$  for  the $B^{-}_c \rightarrow \bar{\nu}_{l} l^- X(3930)$ decay. The result is depicted as a function of $ M_{inv} (D^{*} \bar{D}^{*})$ in Fig. \ref{fig:J2}. The dash curve in this figure is the phase space. The difference with phace space is also apparent.

\begin{center}
  \begin{figure}
  \vspace{10 mm}
   \centering
  \resizebox{0.5\textwidth}{!}{
  \includegraphics{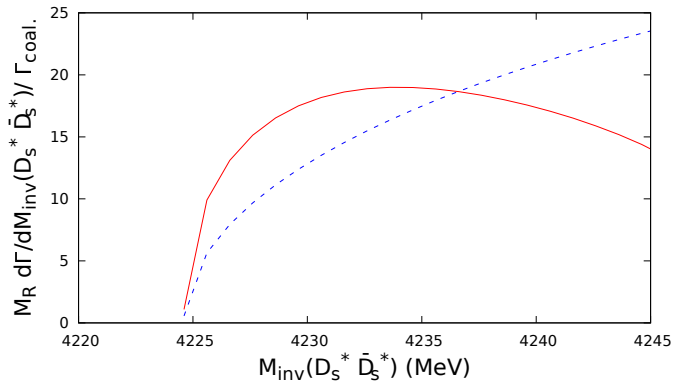}  }
  \caption{ The  $   \frac{M_{R}}{\Gamma_{\rm coal}}\frac{d \Gamma_{i}}{dM_{\rm inv}}  $ as a function of $ M_{inv} ( D^{*}_{s} \bar{D}^{*}_{s})$ for the $B_c \rightarrow \nu_{l} l^+ X(4160)$  decay. The dashed line corresponds to  phase space.}
   \vspace{10 mm}
  \label{fig:J2pr}
  \end{figure}
  \end{center}

  Finally, we show the result for the $B^{-}_c \rightarrow \bar{\nu}_{l} l^- X(4160)$  decay in Fig. \ref{fig:J2pr} as a function of the $  D^{*}_{s} \bar{D}^{*}_{s} $ mass distribution. We observe in this case that the mass distribution close to the $  D^{*}_{s} \bar{D}^{*}_{s} $ is quite different from the phace space.
  
   As important as the shape, showing the presence of a resonance below threshold, the values in the scale, corresponding to ratios, are absolute values of our predictions, tied to the molecular nature of these resonances and their strong coupling to $ D^{*} \bar{D}^{*} $ and $  D^{*}_{s} \bar{D}^{*}_{s} $. We should note that we have compared the $D^{*} \bar{D}^{*}$ or  $  D^{*}_{s} \bar{D}^{*}_{s} $ production with the production of each particular resonance. This implies that in the experiment the $S $-wave is separated from the $ D $-wave in each case, something that is at reach in present partial wave analysis of data \cite{Aaij:2016nsc, Aaij:2015tga}

\section{Conclusions}

We have studied the semileptonic decay of $ B_{c} $ in the reaction  $B^{-}_c \rightarrow \bar{\nu}_{e} e^- X$, with $ X $ any of the resonances $X(3930)~(2^{++})$, $X(3940)~(0^{++})$ and  $X(4160)~(2^{++})$. The main point of the approach is that we treat these resonances as dinamically generated from the vector-vector interaction in the charm sector.  

The $X(3940)$ and $X(3930)$ states are basically $ D^{*} \bar{D}^{*} $ molecules in that approach, although they also couple to other channels with a smaller intensity. To produce these $ X $ states one proceeds in three steps. The first one looks into the elementary process $B^{-}_c \rightarrow \bar{\nu}_{e} e^- c \bar{c}$. In the second step the $ c \bar{c} $ pair hadronizes producing an extra $ \bar{q} q$ with the vacuum quantum numbers, which leads to $ D^{*} \bar{D}^{*} $ and $  D^{*}_{s} \bar{D}^{*}_{s} $ pairs. In the last step these mesons are allowed to undergo final state interaction from where the three resonances appear. By analogy with other reactions producing scalar mesons in the final state, we make an estimation of the rate of $B^{-}_c \rightarrow \bar{\nu}_{e} e^- X(3940)$. For the production of the two tensor states we can not obtain the absolute rate of production, but we can obtain the ratio for the $X(3930)$ and $X(4160)$ states. We also look at the production of  $ D^{*} \bar{D}^{*} $ and $  D^{*}_{s} \bar{D}^{*}_{s} $ close to threshold and we can make predictions of the ratio of this differential mass distribution to the rate of resonance production, which are tied to the nature of these resonances as dynamically generated from the vector-vector interaction in the charm sector. As more decay modes of $ B_{c} $ become available, it would be interesting to look into these modes which will provide good information on the nature of these resonances.

\section*{Acknowledgments}

One of us, N. I., wishes to acknowledge
the support by Open Partnership Joint Projects of JSPS Bilateral Joint Research Projects.
This work is partly supported by 
the Grants-in-Aid for Scientific Research No.15H06413,
the Spanish Ministerio de Economia y
Competitividad and European FEDER funds under the contract number
FIS2011-28853-C02-01 and FIS2011-28853-C02-02, and the Generalitat
Valenciana in the program Prometeo II-2014/068.

\bibliographystyle{plain}

\end{document}